\documentclass[10pt]{article}

\usepackage{fullpage}
\usepackage{setspace}
\usepackage{parskip}
\usepackage{titlesec}
\usepackage[section]{placeins}
\usepackage{xcolor}
\usepackage{breakcites}
\usepackage{lineno}
\usepackage{hyphenat}

\PassOptionsToPackage{hyphens}{url}
\usepackage[colorlinks = true,
            linkcolor = blue,
            urlcolor  = blue,
            citecolor = blue,
            anchorcolor = blue]{hyperref}
\usepackage{etoolbox}
\makeatletter
\ifdefined\H@old@part
  \apptocmd{\H@old@part}{\gdef\cont@name@part{#2}}{}{\cont@err{\H@old@part}}
\else
  \apptocmd{\@part}{\gdef\cont@name@part{#2}}{}{\cont@err{\@part}}
\fi
\makeatother

\usepackage[square,sort,comma,numbers]{natbib}

\renewenvironment{abstract}
  {{\bfseries\noindent{\abstractname}\par\nobreak}\footnotesize}
  {\bigskip}

\titlespacing{\section}{0pt}{*3}{*1}
\titlespacing{\subsection}{0pt}{*2}{*0.5}
\titlespacing{\subsubsection}{0pt}{*1.5}{0pt}

\usepackage{authblk}

\usepackage{graphicx}
\usepackage[space]{grffile}
\usepackage{latexsym}
\usepackage{textcomp}
\usepackage{longtable}
\usepackage{tabulary}
\usepackage{booktabs,array,multirow}
\usepackage{amsfonts,amsmath,amssymb}
\providecommand\citet{\cite}
\providecommand\citep{\cite}

\newif\iflatexml\latexmlfalse

\AtBeginDocument{\DeclareGraphicsExtensions{.pdf,.PDF,.eps,.EPS,.png,.PNG,.tif,.TIF,.jpg,.JPG,.jpeg,.JPEG}}

\usepackage[utf8]{inputenc}
\usepackage[english]{babel}

\usepackage{float}

\usepackage[margin=1.5in]{geometry}

\begin{document}

\title{Accurate Diagnosis of Cancer by Machine Learning Classification of the Whole Genome Sequencing Data}

\author[*]{Arash Hooshmand}%
\affil[*]{KTH Royal Institute of Technology}%

\vspace{-1em}

  \date{}

\begingroup
\let\center\flushleft
\let\endcenter\endflushleft
\maketitle
\endgroup

\selectlanguage{english}
\begin{abstract}
Supervised machine learning can precisely identify different cancer tumors at any stage by classifying cancerous and healthy samples based on their genomic profile. We have developed novel methods of MLAC (Machine Learning Against Cancer) achieving perfect results with perfect precision, sensitivity and specificity. We have used the whole genome sequencing data acquired by next generation RNA sequencing techniques in The Cancer Genome Atlas and Genotype-Tissue Expression projects for cancerous and healthy tissues respectively. Indeed, a creative way to work with data and general algorithms has resulted in perfect classification i.e. all precision, sensitivity and specificity are equal to 1 for most of different tumor types even with modest amount of data. Our system can be used in practice because once the classifier is trained, it can be used to classify any new sample of new potential patients. One advantage of our work is that the aforementioned perfect precision and recall are obtained on samples of all stages including very early stages of cancer; therefore, it is a promising tool for diagnosis of cancers in early stages. Another advantage of our novel model is that it works with normalized values of RNA sequencing data, hence people's private sensitive medical data will remain hidden, protected and safe. This type of analysis will be widespread and economical in the future and people can even learn to receive their RNA sequencing data and do their own preliminary cancer studies themselves which has the potential to help the healthcare systems.%
\end{abstract}%

\par\null

\section*{1. Introduction}

{\label{252565}}

Cancer is one of the most common risk factors that threatens people's lives and is still a severe unsolved problem as one of the main leading causes of death \cite{b13}. Early diagnosis plays a vital role in cancer treatment and survival. There are dozens of different types of cancers affecting different organs of body because cancer can start at any part of the body. \cite{b14} In fact, cancer starts when cells in the body begin to grow out of control and is usually caused by genetic mutations in different cells. Yet the main underlying reasons that cause these mutations are unknown. In recent years, along with generation of big data by high throughput omics technologies, applications of ML (Machine Learning) in diagnosis, treatment and prognosis of cancers are hot topics of research. Consequently, computers have turned out to be promising tools and reliable assistants to contribute in new discoveries based on analysis of the big data generated by high throughput technologies. In this work, we have proposed a novel approach to use genetic transcriptomic data leading to great results with perfectly accurate distinguish between WES (Whole Exome Sequencing) RNA profiles of 22 different main cancers from TCGA (The Cancer Genome Atlas), \cite{b15} and their corresponding healthy tissue samples from GTEx (Genotype-Tissue Expression) project \cite{b16} with samples from different numbers of people as illustrated in Table 1. 

Table 1 also contains reported numbers of estimated new cases and estimated deaths because of each cancer out of estimated 1806590 new cases of all cancers with estimated 606520 new deaths in 2020 in the U.S. for instance.\cite{b1} Reusability and transfer learning are among the main advantageous of ML that means once a machine is well-trained and could distinguish cancerous from noncancerous tissues, it can be fed by new data of new samples acquired from new people and the new sample will be classified correctly with a high likelihood. Therefore, the utilization of ML systems that can detect cancerous genome even at the earliest stages by NGS (Next Generation Sequencing) technology is likely to be a killer application.  

{\label{350486}}

\begin{table}[ht]
\centering
\begin{tabular}{|l|l|l|l|l|l|l|}
\hline
TCGA abbr. & Cancer (TCGA) & Organ (Gtex) & \#Cancer & \#Healthy & New Cases & Deaths \\
\hline
ACC & Adrenocortical Carcinoma & Adrenal Gland & 77 & 128 & - & -  \\
\hline
BLCA & Bladder Carcinoma & Bladder & 407 & 9 & 81400 & 17980  \\
\hline
LGG & Lower Grade Glioma & Brain (Nerves) & 508 & 1152 & 23890 &  18020  \\
\hline
BRCA & Breast invasive Carcinoma & Breast & 1091 & 179 & 279100 & 42690  \\
\hline
CESC & Cervical cancers & Cervix & 304 & 10 & 13800 & 4290  \\
\hline
LAML & Acute Myeloid Leukemia & Bone Marrow & 173 & 70  & 6150 & 1520  \\
\hline
COAD & Colon Adenocarcinoma & Colon & 286 & 308  &  104610 & 53200  \\
\hline
ESCA & Esophageal Carcinoma & Esophagus & 181 & 653  & 18440 & 16170  \\
\hline
GBM	& Glioblastoma Multiforme &  Brain & 518 & 1152 & incl. Brain & @ LGG \\
\hline
KIRC & Kidney Renal Clear Cell & Kidney (Pelvis) & 530 & 28  & 73750 & 14830  \\
\hline
LIHC & Liver Hepatocell. Carci. & Liver (\& bile duct) & 369 & 110    & 42810 & 30160 \\
\hline
LUAD  & Lung Adenocarcinoma & Lung (\& bronchus) & 513 & 288 &    228820 & 135720 \\
\hline
LUSC & Lung Squamous Cell & Lung (\& bronchus) & 498 & 288 &  incl. Lung & @LUAD \\
\hline
OV & Ovarian cystadenocarci. & Ovary (Serous)  & 420 &  88       &	21750 & 13940 \\
\hline
PAAD & Pancreatic Adenocarci. &  Pancreas & 178&  167               & 57600 & 47050 \\
\hline
PRAD & Prostate Adenocarci. & Prostate & 495 &  100  &  191930 & 33330  \\
\hline
READ & Rectum Adenocarci. &  Colon & 91 & 308 &   43340 &  @COAD \\
\hline
SKCM  & Skin Cutaneous Melanoma  &  Skin  & 102 & 812  &  108420 & 11480 \\
\hline
STAD & Stomach Adenocarcinoma & Stomach  & 414 & 174   &  27600 & 11010 \\
\hline
TGCT & Testicular Germ Cell Tumors & Testes & 132 & 165         &  9610 & 440 \\
\hline
THCA & Thyroid Carcinoma & Thyroid & 504 & 279   & 52890 & 2180 \\
\hline
UCEC & Uterine Endometrial Carci. & Uterus (Corpus) &  180 & 78        & 65620 & 12590 \\
\hline
\end{tabular}
\caption{\label{tab:example1} Number of samples of cancerous and healthy tissues obtained from TCGA and Gtex plus estimated number of new cases and deaths of corresponding cancer in 2020 in the US.}
\end{table}

\section*{2. Methods}

{\label{350277}}

ML and AI (Artificial Intelligence) is rapidly opening their positions in medical and pharmaceutical sciences. Different models of ML have been tested successfully in recent years in many projects as well as in this work and have returned decent results. Naïve Bayes, Support Vector Machines, Decision Trees, Random Forest, Logistic Regression and K Nearest Neighbors are examples of general supervised ML algorithms that have reportedly given great results in different projects in different fields of science and are analyzed in our project too. In addition to them, an unsupervised ML method i.e. K-Means is also tested. In this work we came up with a practical approach of applying ML for cancer diagnosis that is effective and robust in different ML algorithms we have tried. Since they are most well-known common ML algorithms, we will briefly introduce them in the following paragraphs. 
On the other hand, the WES genetic information obtained by NGS are openly available on TCGA, Gtex and other online public databases. However, we do not review the technical details of RNA sequencing techniques because they are out of scope of the current paper. The focus of our work in this article is to receive the data from two aforementioned open databases, train the ML classifiers with them, and validate them. To do it, we have used the following ML techniques: %rather than the tasks done prior to this procedure and potentially will be necessary in the future during the treatments based on these new techniques.

\subsection*{2.1 Naive Bayesian classifiers}
\label{sec:1}
Bayes' theorem was proposed by the English Thomas Bayes in 1763 when he was trying to prove the existence of God by means of statistical inference. \cite{b7} Bayesian statistics are used in estimates based on anticipated subjective knowledge. Therefore, the implementations of this theorem adapt with use and allow combining the fusion of data from two or more different sources and expressing them in terms of likelihood. Naive Bayesian Classifier is an implementation of Bayes' theorem, with some additional simplifying hypotheses, which allow applying an independence hypothesis, between the predictor variables, hence "Naive" is added to the name of these implementations because a naive Bayesian classifier assumes that the features of a class / object are not related to each other i.e. the presence of a particular feature is not related to the presence or absence of another. In this way each feature independently contributes to the probability of a given class. In return,  Bayes Classifiers can easily be trained, require little data to train, and can classify big data quickly. Despite the fact that naive Bayes classifiers are amazingly simple, they have worked quite well in many real-world situations, including our cancerous/healthy tissue classification. Our naive Bayes classifier requires a small amount of training data and is fast and accurate as reflected in the Results section. On the flip side, although naive Bayes is known as a decent classifier, it is known to be a bad estimator in a sense that one cannot rely on its parameters for extraction of feature importance. \cite{b9} 

\subsubsection*{2.1.1 Bayesian model function}
\label{sec:2}
More formally, as shown by equations 1-6, Bayesian classifiers are, indeed, probabilistic classifiers using Bayes rule i.e. \begin{equation} P(A \mid B) = P(A) P(B \mid A)/P(B) \end{equation} For example, A can be the prior probability of cancer and B the posterior probability of cancer; given positive cancer test result is the product of the prior times the sensitivity i.e. the chance of a positive result given cancer. Indeed, a naive Bayesian classifier accomplishes statistical inference based on maximum likelihood estimation i.e. setting the parameters of the probability distribution in a way that maximises the goodness of fit of a statistical model to the training data via joint probability distributions of the training samples. In technical words, the likelihood function describes a hyper surface whose peak, if it exists, is an arrangement of model parameters values and coefficients that maximize the probability of drawing the obtained sample. \cite{b8} In its more general form, according to Sci-kit Learn website documentation, Bayes’ theorem states the following relationship, given class variable $y$ and dependent feature vector $x_1$ through $x_n$:
\begin{equation} P(y \mid x_1, \dots, x_n) = \frac{P(y) P(x_1, \dots, x_n \mid y)}
                                 {P(x_1, \dots, x_n)} \end{equation}
Using the naive conditional independence assumption that
\begin{equation} P(x_i | y, x_1, \dots, x_{i-1}, x_{i+1}, \dots, x_n) = P(x_i | y) \end{equation}
for all i, this relationship is simplified to
\begin{equation} P(y \mid x_1, \dots, x_n) = \frac{P(y) \prod_{i=1}^{n} P(x_i \mid y)}
                                 {P(x_1, \dots, x_n)} \end{equation}
Since $P(x_1, x_2,..., x_n)$ is constant given the input, we can use the following classification rule: 
 \begin{equation} P(y \mid x_1, \dots, x_n) \propto P(y) \prod_{i=1}^{n} P(x_i \mid y) => \\\hat{y} = \arg\max_y P(y) \prod_{i=1}^{n} P(x_i \mid y) \end{equation} We can use MAP (Maximum A Posteriori) estimation to estimate $P(y)$ and $P(x_i \mid y)$; the former is then the relative frequency of class  in the training set. The different naive Bayes classifiers differ mainly by the assumptions they make regarding the distribution of $P(x_i \mid y)$. \cite{b9}
 GaussianNB implements the Gaussian Naive Bayes algorithm for classification. Thus, the likelihood of the features is assumed to be Gaussian: \begin{equation}P(x_i \mid y) = \frac{1}{\sqrt{2\pi\sigma^2_y}} \exp\left(-\frac{(x_i - \mu_y)^2}{2\sigma^2_y}\right) \end{equation} where the parameters $\sigma_y$ and $\mu_y$ are estimated using maximum likelihood.

\subsection*{2.2 Support vector machines}
Support Vector Machines (SVMs) are a set of supervised learning algorithms developed by Vladimir Vapnik and his team at AT\&T Labs. \cite{b17,b18,b19} These methods are related to both classification and regression problems. In classification, given a set of sample training examples, we can label the classes and train an SVM to build a model that predicts the class of a new sample. Intuitively, an SVM is a model that represents the sample points in space, separating the classes into spaces as distant as possible using a separation hyperplane defined as the vector between the two points, of the two classes, closest to the which is called the support vector. When the new samples are put before the model, depending on the spaces to which they belong, they can be classified into the right class.

\begin{figure}[ht]
\centering
\includegraphics[width=\linewidth]{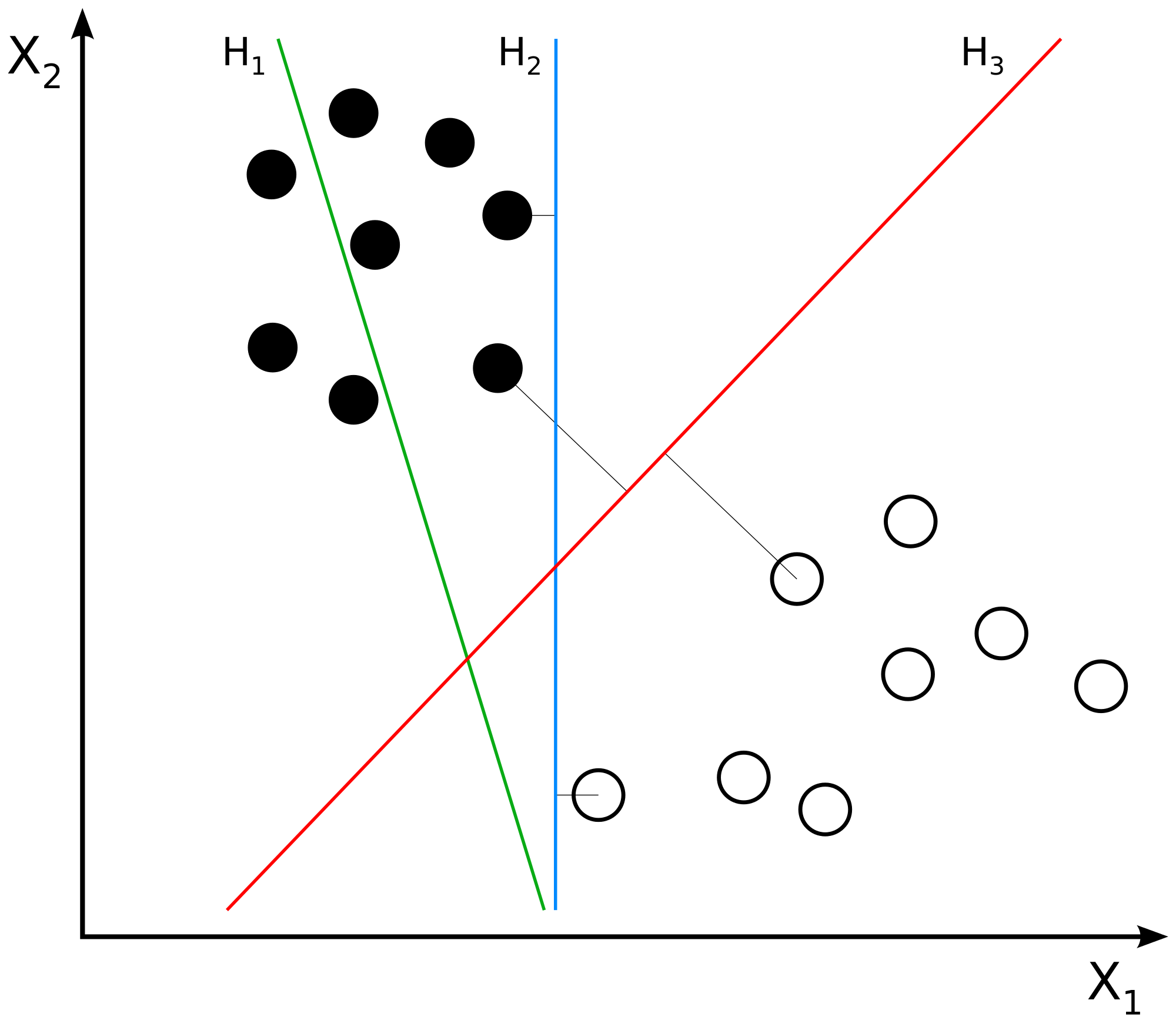}
\caption{Svm separating hyperplanes, from article "Support vector machine". In Wikipedia (2012). Accessed June 27, 2020.}

\label{fig:SVM}
\end{figure}

\subsubsection*{2.2.1 SVM Model function}
More formally, an SVM builds a set of hyperplanes \cite{b20} in a very high (or even infinite) dimensional space that can be used in classification or regression problems. A good separation between the classes will allow a correct classification. In this concept of optimal separation is where the fundamental characteristic of SVM resides: this type of algorithms search for the hyperplane that has the maximum distance (margin) with the points that are the closest to it. This is why SVMs are also sometimes referred to as maximum margin classifiers. In this way, the points of the vector that are labeled with one category will be on one side of the hyperplane and the cases that are in the other category will be on the other side. SVM algorithms intrinsically belong to the family of linear classifiers. The vector formed by the points closest to the hyperplane is called the support vector. Using tricks such as kernel functions, SVMs can also be an alternative training method for polynomial classifiers, radial base functions, and multilayer perceptron neural networks. Figure 1 illustrates SVM mechanism.

\subsection*{2.3 Decision trees}
A decision tree is a tree-like map of the possible outcomes of a series of decisions and only contains conditional control statements comparing possible actions with each other according to their costs, probabilities and utilities. The goal of a decision tree is to break down all the available visit data that a system can learn from and group it so that each group's visits are as similar to each other as possible with respect to the goal metric. Between groups, however, visits are as different as possible relative to the goal metric (for example, conversion rate). The decision tree takes into account the different variables existing in the training set to determine how to divide the data MECE (mutually exclusive, collectively exhaustive) into these groups or leaves to maximize the goal. A decision tree typically starts with a single node and then branches out into possible outcomes. Each of the outcome nodes creates additional nodes, which branch into other different possibilities. This creates a structure similar to that of a tree. There are three different types of nodes: probability nodes, decision nodes, and terminal nodes. A chance node,typically represented by a circle, shows the probabilities of certain outcomes. A decision node,typically represented by a square, shows a decision to be made, and a terminal node,typically represented by a triangle, shows the final result of a decision route. Decision trees are still popular for advantages such as requiring minimal data processing and being easily understood, updated, (new options can be added to existing trees), and integrated  with other decision-making tools. However, decision trees can become very complex. In those cases, a more compact influence diagram can be a good alternative focusing on fundamental goals, inputs, and decisions. \cite{b21}

\subsection*{2.4 Random forests}
By iteratively applying the algorithm that creates decision trees with different parameters on the same data, we get what is called a random forest. This algorithm is one of the most efficient prediction methods for big data, since it averages the performance of many different models with noise and impartially reduces the final variability of the set. In reality, what is done is to build different training and test sets on the same data, which generates different decision trees on the same data. The union of these trees of different complexities and with data of different origin, although from the same set, results in a fairly stable random forest whose main characteristic is that it creates more robust models than what could be obtained by creating a single decision tree on the same data. In classification, the class that is the mode of classes will be output. \cite{b22,b23}

\subsection*{2.5 Binary logistic regression}
Logistic regression is a group of statistical techniques that aim to test hypotheses or causal relationships when the dependent variable is nominal. 
Despite its name, it is not an algorithm applied in regression problems, in which continuous values are dealt with, but it is a method for classification problems, in which a binary value i.e. either 0 or 1 is obtained. For example, a classification problem is to identify if a given tumor is malignant or benign. With the logistic regression, the relationship between the dependent variable i.e. the statement to be predicted, with one or more independent variables i.e. the set of features available for the model is determined. To do this, it uses a logistic function that determines the probability of the dependent variable. As previously mentioned, what is sought in these problems is a classification, so the probability must be translated into binary values for which a threshold value is used. If the probability values were above the threshold value, the statement is true and vice versa. Generally this value is 0.5, although it can be increased or decreased to manage the number of false positives or false negatives. \cite{b24}

\begin{figure}[ht]
\centering
\includegraphics[width=\linewidth]{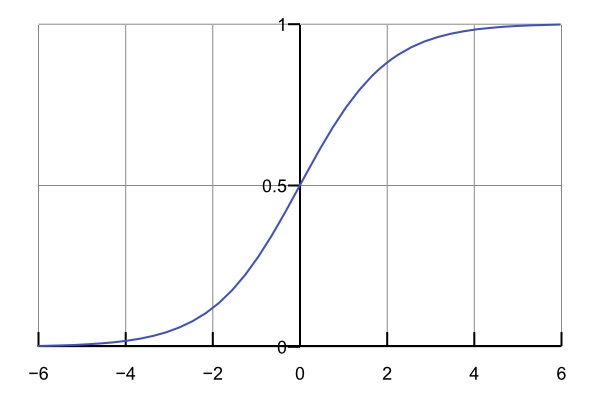}
\caption{Sigmoid function as the logistic curve}

\label{fig:Sigmoid}
\end{figure}

The function that relates the dependent variable to the independent ones is also usually either the sigmoid function or a function similar to it such as tanh and softmax. The sigmoid function is an S-shaped curve that can take any value between 0 and 1, but never values outside these limits. The equation that defines the sigmoid function is  $f (x) = 1/(1+e^{- x}) $ where x is a real number. In the equation you can see that when x tends to minus infinity the function tends to zero. On the other hand, when x tends to infinity the function tends to unity. Figure 2 shows a graphical representation of the logistic function (sigmoid function).
Logistic regression is a technique widely used because of its effectiveness and simplicity. As one of its advantages, it is not necessary to have large computational resources, neither in training nor in execution. Furthermore, the results are highly interpretable that is one of its main advantages. The weight of each of the features determines the importance it has in the final decision. Therefore, it can be affirmed that the model has made one decision or another based on the existence of one or another certain feature. What in many applications is highly desired in addition to the model itself. Regarding its disadvantages is the impossibility of directly solving non-linear problems because the expression that makes the decision is linear. For example, in the event that the probability of a class is initially reduced with a feature and subsequently increased, it cannot be registered with a logistic model directly. If necessary, this feature should previously be transformed so that the model can record this non-linear behavior. In these cases, it is better to use other models such as decision trees. Indeed, the important point is that the target variable must be linearly separable. Otherwise, the logistic regression model will not classify correctly. In other words, there must be two "regions" with a linear border in the data. Another drawback is the dependency it shows on the features. Logistic regression is not one of the most powerful algorithms that exist. It would easily be surpassed by other more complex classifiers. Finally, in machine learning, there are classifiers that can work with multiple classes, such as Decision Trees or Random Forest. On the other hand, there are others that do not, such as Logistic Regression. However, it is always possible to use tricks to use logistic regression in classification problems with multiple classes such as: \newline
•	OvA (One versus all): In this strategy, you have to train as many binary classifiers as possible with respect to the classes that are there in the data set. Each of the models predicts the probability that the record belongs to a class. When making a prediction, all classifiers are run and the one with the highest probability is selected. \newline
•	OvO (One versus one): In this strategy, as many models are created as there are pairs of possible outcomes, that is, they have to be trained $(N ^ 2 -N) / 2$ models, where N is the number of possible classes. It means that a classifier will decide only between two possible outcomes. As in the previous case, when making a prediction, all classifiers are run and the one with the highest probability is selected.

\subsection*{2.6 K Nearest Neighbors}
K-Nearest-Neighbor is a simple nonparametric instance-based algorithm of supervised ML. It can be used to classify new samples (discrete values) or to predict (regression, continuous values). It is essentially used to classify values by searching for the most similar (by proximity) data points learned in the training stage and making guesses of new points based on the prior classification. \cite{b25,b26}
Unlike K-means, which is an unsupervised algorithm where the "K" means the number of groups that we want to classify, in K-Nearest Neighbor the "K" means the number of "neighboring points" that we consider in the vicinity to classify the "n" groups which are already known in advance. It is a method that simply searches the closest observations to the the point of interest that is to be classified and classifies it based on most of the data that surrounds it. As we said before, K nearest neighbor algorithm is: \newline
•	Supervised: that means that we have tagged our training data set, with the class or expected results. \newline
•	Instance-based: that means that our algorithm does not explicitly learn a model (such as in Logistic Regression or Decision Trees). Instead, it memorizes the training instances that are used as the knowledge-base for the prediction phase. \newline 
KNN is easy to learn and implement. However, it uses the entire data set to train each point and therefore requires a lot of memory and processing resources. For these reasons KNN tends to work best on small data sets and without a huge number of features. To classify the inputs by means of KNN one should: \newline
1.	Calculate the distance between the item to classify and the other items in the training data set.
2.	Select the closest "K" elements (with less distance, depending on the function used). \newline
3.	Carry out a majority vote between K points: those of a class / label that will be determinant in making the final decision. \newline
Taking point 3 into account, we will see that in order to decide the class of a point, the value of K is very important because it defines which are the points their majority will define the group each new point belongs to, and it is especially critical when the new points fall in the borders between groups.

\subsection*{2.7 K-Means}
K-Means is an unsupervised ML algorithm for clustering. It is used when we have a lot of untagged data. The objective of this algorithm is to find K groups (clusters) among the raw data. The algorithm works iteratively to assign each input such as genome sample to one of the K groups based on its features, in here genes. It means that the inputs are grouped based on the similarity of their features. \cite{b27} As a result of executing the algorithm: \newline
•	The centroids i.e. geometric centers of groups will be coordinates of the corresponding K clusters and will be used to label new samples. \newline
•	Labels for the training data set: each tag belonging to one of the K defined groups. \newline

The groups are defined dynamically i.e. their position is adjusted in each iteration of the process until the algorithm would converge. Once the centroids are found, they are analyzed to see what their unique features are, compared to those of the other groups. These groups are the labels that the algorithm generates. The Clustering K-means algorithm is one of the most used methods to find hidden groups or theoretically suspected groups on an unlabeled data set. This can serve to confirm or reject some hypotheses that we would have assumed about our data, and it can also help to discover hidden relationships between data sets. Once the algorithm has executed and obtained the labels, it will be easy to classify new values or samples among the obtained groups. This algorithm works by pre-selecting a value of K. To find the number of clusters in the data, we must run the algorithm for a range of K values, see the results and compare characteristics of the groups obtained. In general, currently there is no exact way to determine the K value, but it can be estimated with acceptable precision using the following technique: One of the metrics used to compare results is the average distance between the data points and their centroid. As long as the value of the mean will decrease as we increase the value of K, we will continue increasing it. The mean distance to the centroid is considered as a function of K and the goal is to find the elbow point where the rate of descent sharpens.

In ML supervised classification methods as well as in K-Means unsupervised clustering algorithm, the input data (the samples) are viewed as a p-dimensional vector (an array or ordered list of p numbers where p in this project is 19627). Then the classifiers based on their criteria distinguish among different groups formed by close/similar samples; e.g. in the Bayesian classifiers, the classifier looks for a hypersurface that maximizes the likelihood of drawing the sample, or in SVMs, it looks for a hyperplane that optimally separates the points of one class from the other, which eventually could have been previously projected to a higher dimensional space. There has been wrong perceptions in the ML community preventing potential achievements; for instance, people try to decrease the number of features to avoid "the curse of dimensionality". While the curse of dimensionality may truly happen in some problems, it may not be an issue in other problems such as ours. Deleting features blindly for fear of dimensionality may only result in losing useful information without need. Researchers usually try to reduce by themselves the assumed learning pressure on the machines brought about by highly redundant dimensions and select a subset of features i.e. genes to reduce the number of features and dimensions. \cite{b10,b11,b12} It may have hurt their results. A strength point of our work is that we consider ML as powerful advanced statistics tool doing heavy statistical analyses, that people themselves cannot do. As a result, we gave all the data corresponding to the WES as feature inputs to the ML at once and it returned almost perfect results quickly and precisely. We thought of 19627 different genes not as too many features but as different pixels of a less than 141*141-pixel photo and it was a very light task for the machine to analyze such a low resolution image and it took only seconds to classify the cancerous and noncancerous cells 100\% precisely.

\subsection*{2.8 Model optimization and settings}
\label{sec:4}

We have employed all the classifiers from Scikit-Learn 0.23.1 with their default settings unless mentioned otherwise. For example, Scikit-Learn's Gaussian Naive Bayes classifier, that is a simple classifier, has only two parameters i.e. priors equal to None and var\_smoothing equal to 1e-9 where var\_smoothing is the portion of the largest variance of all features that is added to variances for calculation stability. We did not touch the defaults, but there were exceptions such as SVM in which we changed two default settings: we decided to use "linear" kernel instead of "rbf" that was the default kernel and also we set "PROBABILITY = TRUE" in order to obtain "predict\_proba" that is a useful attribute to calculate and plot the ROC curve but is ignored in default setting when "PROBABILITY = FALSE". Therefore, except these two minor changes at SVM default settings, all models were run with default settings of Scikit-Learn version o.23. As other settings, for most of the cancers, 90\% specific cancer samples were used as the training dataset and the remaining 10\% used as testing dataset chosen by random using train\_test\_split function of Sci-Kit Learn model selection modul with Random Seed equal to zero. The only exceptions were for bladder and cervix for which the number of healthy samples was too low. Therefore, we used 40\% for training and 60\% for testing in bladder and 70\% for training and 30\% for testing in cervix cancer. However, we analyzed the effect of different data allocation plans from 10\% to 90\% for test/validation set and also tried other random seeds and in particular for K Nearest Neighbor algorithm, we also tried it with different K values. The results were not significantly different and discussed more in the following at the Discussion part of this article. We also decided to mostly publish the results achieved by those classifiers that can do the classification perfectly; however, all six classifiers work well and the imperfect ones also return results close to perfect. The models take 19627-genes WES data as input and after a quick and easy model training with no need to data modification, acceptable classification results are obtained and there are at least two classifiers per cancer that could distinguish both cancerous and healthy tissues perfectly with no error.

\subsection*{2.9 Model evaluation}
\label{sec:5}
Model evaluation produces measures to approximate a classifier's reliability. To distinguish between cancerous and noncancerous cells, since it is a binary classification, we use accuracy, precision, specificity, sensitivity, f1 score, several averaging techniques and ROC curve to evaluate the model. We, indeed, use Sci-kit Learn Metrics Classification Report that returns precision, recall and f1 score for each of two classes. In binary classification, recall of the positive class is called “sensitivity”; and recall of the negative class is “specificity”. In what follows, the principal terms and then equations7-22 derivations based on confusion matrix such as accuracy, specificity, sensitivity, f1 score are given to review and compare: \newline
•	Condition positive (P): the number of real positive cases in the data \newline
•	Condition negative (N): the number of real negative cases in the data \newline
•	True positive (TP) or hit \newline
•	True negative (TN) or correct rejection \newline
•	False positive (FP), false alarm or type I error \newline
•	False negative (FN), miss or type II error \newline
 \newline
Sensitivity, recall, hit rate, or true positive rate (TPR): % \newline 
\begin{equation} TPR = TP / P = TP / (TP + FN) = 1 - FNR \end{equation} %\newline
% \newline
Specificity, selectivity or true negative rate (TNR): % \newline 
\begin{equation} TNR = TN / N = TN / (TN + FP) = 1 - FPR \end{equation} % \newline
% \newline
Precision or positive predictive value (PPV) is the ratio of the correctly labeled samples by our program to all labeled ones in reality. 
\begin{equation} PPV = TP / (TP + FP) = 1 - FDR \end{equation} % \newline
Precision can be calculated only for the positive class i.e. class 1 that shows cancer or can be evaluated for each one of the two classes independently treating each class as it is the positive class at time, and the latter is done in Sci-kit Learn Metrics Classification Report as shown in table 1. \newline
 \newline
Negative predictive value (NPV): % \newline 
\begin{equation} NPV = TN / (TN + FN) = 1 - FOR \end{equation} % \newline
% \newline
Miss rate or false negative rate (FNR): % \newline 
\begin{equation} FNR = FN / P = FN / (FN + TP) = 1 - TPR \end{equation} % \newline
% \newline
Fall-out or false positive rate (FPR): % \newline 
\begin{equation} FPR = FP / N = FP / (FP + TN) = 1 - TNR \end{equation} % \newline
% \newline
False discovery rate (FDR): % \newline 
\begin{equation} FDR = FP / (FP + TP) = 1 - PPV \end{equation} % \newline
% \newline
False omission rate (FOR): % \newline 
\begin{equation} FOR = FN / (FN + TN) = 1 - NPV \end{equation} % \newline
% \newline
Accuracy (ACC): % \newline 
\begin{equation} ACC = (TP + TN) / (T + N) = (TP + TN) / (TP + TN + FP + FN) \end{equation} % \newline
% \newline
The harmonic mean of precision and sensitivity or f1-score (F1): % \newline 
\begin{equation} F1 = 2.PPV.TPR / (PPV + TPR) = 2.TP / (2.TP + FP + FN) \end{equation} % \newline 
% \newline
Since we are using Sci-kit Learn Metrics Classification Report to show the results as shown in table 1, we also describe the meaning of micro avg, macro avg and weighted avg. used in the report: \newline
\newline
Micro-average of precision (MIAP):
\begin{equation} MIAP = (TP1+TP2)/(TP1+TP2+FP1+FP2) \end{equation} % \newline
Micro-average of recall (MIAR):
\begin{equation} MIAR = (TP1+TP2)/(TP1+TP2+FN1+FN2) \end{equation} % \newline
Micro-average of f-Score (MIAF) would be the harmonic mean of the two numbers above. % \newline
\begin{equation} MIAF = 2.MIAP.MIAR / (MIAP + MIAR) \end{equation}  
% \newline
Macro-average of precision (MAAP):
\begin{equation} MAAP= (Precision 1 + Precision 2)/2 \end{equation}  % \newline
Macro-average of recall (MAAR):
\begin{equation} MAAR= (Recall 1 + Recall 2)/2 \end{equation} % \newline
Macro-average of f-Score (MAAF) would be the harmonic mean of the two numbers above. % \newline
\begin{equation} MAAF = 2.MAAP.MAAR / (MAAP + MAAR) \end{equation} 
%\newline
Macro-average is suitable to know how the system performs overall across different sets of data but should not be considered in any specific decision-making because it calculates metrics for each label and finds their unweighted mean i.e. it does not take label imbalance into account, while in our case, the labels are highly imbalanced in many sets e.g. 1091 vs. 179. On the other hand, micro-average is a useful tool and returns measures for our decision-makings especially when coupled healthy-cancerous datasets vary in size because it calculates metrics globally by counts the total true positives, false negatives and false positives. Finally, Weighted-average, according to Sci-kit Learn documentation on f1-score metrics, calculates metrics for each label, and finds their average weighted by support (the number of true instances for each label). This alters "macro" to account for label imbalance; consequently, it can result in an F-score that is not between precision and recall.

The ROC (Receiver Operating Characteristic) curve is created by plotting the true positive rate (TPR) or sensitivity against the false positive rate (FPR) i.e. (1-specificity) at different threshold settings. Varying the decision threshold from its maximal to its minimal value results in a piecewise linear curve from (0,0) to (1,1), such that each segment has a non-negative slope (Figure 3). This ROC curve is the main tool used in ROC analysis and in general, can be used to address a range of problems; however, in our illustrated case where the performance is perfect, it is just a visual endorsement for the perfect classification and the corresponding AUC (Area Under the ROC Curve) is its maximum i.e. 1.

\par\null

\begin{table}[ht]
\centering
\begin{tabular}{|l|l|l|}
\hline
Confusion matrix & Predicted 0 & Predicted 1 \\
\hline
Class 0 & TN = 1 & FP = 0  \\
\hline
Class 1 & FN = 0 & TP = 1  \\
\hline
\end{tabular}
\caption{\label{tab:example2} Typical confusion matrix with no confusion for perfect identification of cancerous tumors by either of GNB, SVM, DCT, RFC, LGR and KNN(K=3) where sensitivity, specificity and precision are all 100\%.}
\end{table}

\begin{figure}[ht]
\centering
\includegraphics[width=\linewidth]{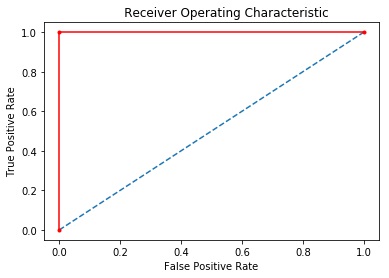}
\caption{Typical ROC curve of perfecrt classification done by classifiers on most of tumor types.}

\label{fig:ROC}
\end{figure}

\section*{3. Results}

{\label{432859}}

The classification performance on a total of 7971 cancerous WES samples of 22 specific cancers from TCGA open public database and their corresponding healthy tissues with 4798 WES samples from the GTEx project were studied in 150+ ML models. Each sample included the normalized volume of 19627 genes as ppm and the data corresponding to both cancerous and healthy samples of each organ are separately and directly fed to the machine to do heavy statistical calculations on the high dimensional data. As result, cancerous samples of all 22 types of tumors at any stages were correctly identified and separated from their corresponding healthy samples. The task was accomplished and compared via six supervised ML classifiers i.e. GNB, SVM, DCT, RFC, LGR and KNN which all showed perfect and in some cases near-perfect performance as shown in Table 3. In addition, Table 3 contains the results for K-Means as an unsupervised clustering technique that was applied to evaluate the ability of the algorithm to distinguish between cancerous and noncancerous cells of different organs as its two main clusters. Clustering, in general, is different from classification and its algorithms such as K-Means have their own evaluation techniques. However, we have applied a trick by using classification accuracy to see how much K-Means clustering to two clusters matches the difference of our interest i.e. two classes of healthy and cancerous cells of an organ or two classes of two different cancers. The results were impressive in here too. In some cases, clustering matched 100\% with our class labels as healthy and cancerous. One tricky point in interpretation of results is that unlike classification in which the higher the accuracy, the better; in clustering 50\% accuracy is the worst while very high and very low measurements of accuracy are equivalently great because, for example, an accuracy equal to zero in clustering means that the clustering algorithm, in here K-Means, has labelled all our class 0 i.e. healthy samples as cluster 1, and all our class 1 labels i.e. cancers as cluster 0. Therefore, any percentage of accuracy and its counterpart i.e. 100 minus that percentage are equally good while 50\% shows maximum entropy and least match with our classes and labels. The clustering performance was impressive in several cases, especially that cancerous and noncancerous cells of Pancreas and Testis were 100\% accurately separated into two distinguished clusters as shown in Table 3. There were also models successfully employed to further distinguish between different types of an organ's cancer and two types of cancer were also separated perfectly. The performance of supervised and unsupervised methods with different parameters were also studied. For most of changes such as different random seeds, allocating different volumes of data for training and testing stages and different values for K in KNN, the consequent changes were negligible as shown in Table 42, Table 43 and Table 44 respectively. The exception was GNB classification performance on LIHC that improved to at leasr 98\% in other settings. However, there are parameters that can change and deteriorate the results. For example, SVMs with linear and rbf kernels are significantly different: while linear kernel accomplish the classification perfectly in 19 out of 22 cancer types with three exceptions that are LGG, COAD and GBM with 99\% of accuracy and also f1-score of 0.99, the results with rbf kernel were not acceptable. An example of LUSC is presented in Table 41 to be compared with Table 42 that includes the results for SVM with linear kernel. The most common typical ROC curve and Confusion Matrix for most of our cancer classifications are illustrated in Figure 3 and Table 2. Therefore, AUC is naturally 1 for most of classifiers in classification of most of cancerous-healthy samples of specific organs. We also analyzed these ML classifiers' capability to distinguish between two types of cancers. Whine no longer as 100\% perfect results as before when they classified between healthy and cancerous tissue, yet the results were great with accuracy and f1-scores above 90\% that is better than the previous works we knew about. The classifiers not only separated samples belonging to two different cancers of brain as well as lung, but also separated classified well tumors of lung's LUSC and Bladder's BLCA which were previously reported to be very similar and confusing for deep neural network classifiers. \cite{b11} For more details refer to Discussions part of this paper below.

\begin{table}[ht]
\centering
\begin{tabular}{|l|l|l|l|l|l|l|l|}
\hline
Accuracy &  GNB & SVM & DCT & RFC & LGR & KNN & KMeans unsupervised clustering \\
\hline
ACC & 1.0 & 1.0 & 0.95 & 1.0 & 1.0 & 1.0 & 0.14=0.86 \\
\hline
BLCA & 1.0 & 1.0 & 0.98 & 1.0 & 1.0 & 1.0 & 0.64 \\
\hline
LGG & 0.98 & 0,99 & 1.0 & 0.99 & 0.99 & 1.0 & 0.77 \\
\hline
BRCA & 1.0 & 1.0 & 0.98 & 1.0 & 1.0 & 0.98 & 0.91 \\
\hline
CECS & 1.0 & 1.0 & 1.0 & 1.0 & 1.0 & 1.0 & 0.69 \\
\hline
LAML & 1.0 & 1.0 & 1.0 & 1.0 & 1.0 & 1.0 & 0.44=0.56 \\
\hline
COAD & 1.0 & 0.99 & 1.0 & 0.99 & 0.99 & 0.97 & 0.18=0.92 \\
\hline
ESCA & 0.99 & 1.0 & 1.0 & 0.99 & 1.0 & 0.98 & 0.59 \\
\hline
GBM & 0.99 & 0.99 & 0.99 & 1.0 & 1.0 & 1.0 & 0.72 \\
\hline
KIRC & 0.98 & 1.0 & 0.98 & 0.98 & 1.0 & 0.98 & 0.77 \\
\hline
LIHC & 0.92 & 1.0 & 0.98 & 1.0 & 1.0 & 0.96 & 0.08=0.92 \\
\hline
LUAD & 1.0 & 1.0 & 0.99 & 1.0 & 1.0 & 0.94 & 0.25=0.75 \\
\hline
LUSC & 1.0 & 1.0 & 0.99 & 1.0 & 1.0 & 0.99 & 0.13=0.87 \\
\hline
OV & 1.0 & 1.0 & 1.0 & 1.0 & 1.0 & 1.0 & 0.69 \\
\hline
PAAD & 1.0 & 1.0 & 1.0 & 1.0 & 1.0 & 1.0 & 1.0 \\
\hline
PRAD & 0.98 & 1.0 & 1.0 & 1.0 & 1.0 & 0.98 & 0.97 \\
\hline
READ & 1.0 & 1.0 & 1.0 & 1.0 & 1.0 & 0.95 & 0.58 \\
\hline
SKCM & 1.0 & 1.0 & 1.0 & 1.0 & 1.0 & 0.99 & 0.65 \\
\hline
STAD & 0.98 & 1.0 & 0.97 & 1.0 & 1.0 & 0.97 & 0.17=0.83 \\
\hline
TGCT & 1.0 & 1.0 & 1.0 & 1.0 & 1.0 & 1.0 & 0.0=1.0 \\
\hline
THCA & 1.0 & 1.0 & 1.0 & 1.0 & 1.0 & 0.95 & 0.30=0.70 \\
\hline
UCEC & 1.0 & 1.0 & 1.0 & 1.0 & 1.0 & 1.0 & 0.69 \\
\hline
\end{tabular}
\caption{\label{tab:example3}ML Classifiers Accuracy in different tumor type classification.}
\end{table}

\section*{4. Discussion}

Our work facilitates effective applications of ML in medical sciences and resulted in excellent classification between cancerous and noncancerous cells of 22 most common cancers. In this work, we did not reduce the dimension of input data and left all the statistical analysis to the ML system and it could do its job very well and perfectly distinguished the cancerous tumors from healthy cells in most of the cancer-classifier combinations, and in the remaining pairs almost perfectly, as shown in Table 3. More details on information summarized in Table 3 can be found in tables 4-45. We learn from this experiment that we are allowed to reduce the dimension only after that the famous problem known as "the curse of dimensionality" has occurred and the system cannot solve it; otherwise, it is better to leave the calculations to the machines and do not blindly reduce the dimensions when there is no issue for machine to deal with all available features and dimensions. For example, in this study, the number of tumor samples ranges from 77 (ACC) to 1,091 (BRCA) and the number of healthy tissue samples ranges from only 9 (Bladder) and 10 (Cervix) to 1152 (Brain) yet our classifiers can learn from the large distance between healthy and cancerous bladder tissues in the 19627 dimensional space and correctly classify all cancerous and healthy tissues very well. For example, after being trained by 6 healthy samples and 287 Cervical squamous cell carcinoma and endocervical adenocarcinoma samples, all classifiers i.e. GNB, SVM, DCT, RFC, LGR and KNN (K=5) could classify the remaining 4 healthy samples and 17 cancerous samples 100\% precisely and only KNN with K=3 failed in some cases with overall F-score of 0.99 instead of 1 because of failure of perfect precision in one class and recall in the other. Meanwhile, the differences among KNN with different values for K were negligible. The important point in here is that, indeed, despite having little data to train, ML classifiers could learn how to classify two groups perfectly thanks to the large number of dimensions and the differences provided with them. If we decrease the number of features/ dimensions while we also lack huge amount of data i.e. a large number of samples, then we may have thrown away useful data and cannot train the system well and it cannot classify that good. It is done by most of researchers before and as a result they get less excellent results. Our hypothesis is that a large number of features and dimensions can compensate lack of large number of samples. Imagine, for example, points in 3-dimension space shown as a Cartesian system. If we have two classes of points that all coordinates of one class are positive numbers and all coordinates of points in the other class are negative numbers, it will be natural that any intelligent system including AI systems can quickly learn to separate the two groups even by a small number of samples. We think that our natural 19K+-dimension space is perfect to distinguish between healthy samples and impaired cancerous samples after watching dozens of samples. Even the results after analyzing a few samples of different classes are amazing.

Another fact to notice is that the samples from TCGA and Gtex are diverse representing different ages, sexes, races and different stages of cancer their statistical details are available on TCGA and Gtex websites. However, TCGA has not provided information on the stage of GBM, LGG, OV, PRAD and UCEC tumors but all other tumors are categorized into four different stages and the classifiers work well on all the data that includes their early satge samples too. Since our classifiers return perfect results on all of them, therefore, the demographic information do not have any serious effect on the performance and we do not need to deal with their statistical details and factors one by one. Another interesting fact is that the problem of separation between cancerous and noncancerous cells seems to be a linear classification problem because SVM with linear kernel can classify almost all cancer types from their corresponding healthy tissue cells perfectly but when SVM was tried with its default kernel i.e. rbf the results were not good. The difference in results of linear and nonlinear separators imply that the samples of two groups are naturally separated linearly.

In comparison, as seen in the result tables in appendix/ supplementary materials summarized in Table 3, SVM with linear kernel, LGR and RFC were almost perfect and slightly superior to other classifiers and return perfect results for almost all tumor types with a few percentage mistake for a few remaining cancers, while KNN and DCT were usually inferior; despite the fact that there were cases that DCT or KNN returned perfectly precise results when others fail to achieve 100\%; or KNN as the worst classifier also has always classified tumors and healthy tissues with no worse than 94\% of accuracy. GNB with our most common settings registered record low accuracy of 92\% only on liver cancer (LIHC) which seems to be its outlier result because with any change, even using less training data i.e. 70\% instead of 90\% its accuracy was always 98\% or more. The simulations are illustrated in the tables below. GNB particularly acts well when there are little data available; therefore, should not be ignored when the resources such as data or computational power are limited.

Finally, we also examined the performance of ML classifiers on several pairs of cancers as shown in the last table i.e. Table 45. First we tried them to classify between two types of tumors of Brain cancer i.e. LGG and GBM, and between two types of Lung cancer i.e. LUAD and LUSC, followed by 2 cancers of Kidney. All of them were separated with accuracy more than 90\% which is amazing because tumors of the same organ are expected to be similar to each other. The lowest rate of correct classification is for distinguish between Colon Adenocarcinoma (COAD) and Rectum Adenocarcinoma (READ) which could be done at the best with about 75\% accuracy by RFC, DFC and KNN which is great because READ and COAD are almost the same things to that extent that Siegel et. al. \cite{b1} have reported the estimated death because of these two cancers together because in practice, many hospitals are confused by them and count READ as COAD as one class of cancer that are the same. Therefore, being able to distinguish between them with more than 70\% accuracy even by K-Means unsupervised clustering is hope-giving. The successful separations between two similar cancers of one organ such as LUAD and LUSC shown in Table 45 may open a new approach in cancer diagnoses i.e. it may be better to first detect if the biopsy sample of an organ such as brain or lung is healthy or malignant tumor, then if it were cancerous, the data can be fed again to the ML classifiers to detect which type of tumor it is. The benefit of this approach is that in binary classifications between healthy and cancerous sample of an organ, there is only one option for cancerous cells. If, for example, the patient suffers from LUSC but the sample is given to a classifier to decide between healthy or LUAD, it will likely be classified as LUAD. Therefore, it has better to either from the beginning classify it as healthy or cancerous including all relevant cancers and then distinguish among cancers or consider the fact and if the sample is recognized as one specific cancer of an organ, then test it again carefully to classify it correctly among potential cancers of the organ or from the beginning use a multi-classifier instead of binary classifiers. For detailed info refer to Tables 20, 21, 22 and 23 and the last row of Table 45.

\section*{5. Conclusion}

{\label{880788}}
These ML systems are trained now and are ready to receive any potential patient's data to recognize if the sampled organ is cancerous or not. It can detect the problem in different stages of cancer accurately; therefore, can be helpful in early diagnosis of cancer. The limitation of our model is that it needs data of samples taken from organs. Yet most of people do not have easy access to this level of their own personal genetic data. Thus the next work can be finding suitable biomarkers in the blood that can detect healthy people and patients only by their blood samples. Furthermore, the world is realizing the importance of creating databases of single cell WESs which will result in more accurate cancer studies and the corresponding big data can also improve ML systems to work perfectly on all cancers hitting each organ of each patient in personalized medicine which can employ most effective treatments for each person based on specific differences. Yet our work is one step forward because the best previous work that, to the best of our knowledge, is done by Sun et. al. \cite{b11} implements complicated deep neural networks that not only is more exhaustive computationally, but also lack clarity as a common feature of deep neural networks while our simple algorithms run quicker, hit their results with better precision and recall, works especially better with little data, and the mechanism of action, hence is not like a black box because the logic of performance is understandable. As an evidence to our claim, we especially tried the classification of a problematic pair of two different tumor types where their deep neural network was confused between them i.e. lung's LUSC and bladder's BLCA. Our classifiers again classified these two tumors with more than 90\% accuracy where their deep neural network classifier was confused between them. \cite{b11} Most notably, using our classifiers is more suitable to obtain important features that play the maximum role in the classification and are the most upregulated or most downregulated genes between cancerous and noncancerous groups. The latter gives invaluable information both directly as well as indirectly to be used along with supportive knowledge of pathways which cause cancers.

\section*{Funding Information}

{\label{974317}}
Thanks to KTH Royal Institute of Technolgy and its Library for their Open Access Publication Grant, as well as Houshmand family and their companies especially GholamAbbas, Atash, Shahab, Shahin and Shadab for their financial support to my studies and projects.

\section*{Research Resources}

{\label{808103}}
The cancerous samples data are from The Cancer Genome Atlas (TCGA) and their corresponding healthy tissue samples are from the Genotype-Tissue Expression (GTEx). All machine learning algorithms are taken from Scikit-learn.

\section*{Acknowledgments}

{\label{749861}}

I'd like to acknowledge all those who have had any contribution to who I am and where I am. Starting from my parents and grand parents, siblings, all my great teachers and everybody who has taught me a single word of science especially those whose knowledge in life sciences or in computer sciences and machine learning has had a direct impact on this research. I also particularly thank Mr. Eng. GholamAbbas Houshmand for his great contribution in financing my studies and projects as well as KTH library for supporting me to publish my paper in open access form to be accessible for all those who are interested in.

\section*{Further Reading}

{\label{153582}}

For readers who may want more information on concepts in your article, please look for the upcoming articles I am going to write as I am developing and testing new ideas and will report them appropriately to the scientific community.

\section*{\texorpdfstring{{Note About
References}}{Note About References}}

{\label{514168}}

References are automatically generated by Authorea.
Select~\textbf{cite~}to find and cite bibliographic resources. The
bibliography will automatically be generated for you in APA format, the
style used by most WIREs titles. If you are writing for~\emph{WIREs
Computational Molecular Science}~ (WCMS), you will need to use
the~Vancouver reference style, so before exporting click
Export-\textgreater{} Options and select a Vancouver export style.~

\selectlanguage{english}
\FloatBarrier

\section*{Author contributions statement}
A.H. is the sole author of this paper and has done all the simulations himself.

\section*{Additional information}

Arash Hooshmand (with ORCiD 0000-0002-9263-0282) is the sole author and hence the corresponding author of this paper and declares that there is no conflict of interests regarding the publication of this paper. Please do not hesitate to contact me via hooshmand@kth.se if you have any questions.

\section*{Appendix I: Supportive Tables}

\begin{table}[ht]
\centering
\begin{tabular}{|l|l|l|l|l|}
\hline
ML / LAML & Precision & Recall & F1-score & Support \\
\hline
Class 0 & 1.00 & 1.00 & 1.00 & 7 \\
\hline
Class 1 & 1.00 & 1.00 & 1.00 & 18 \\
\hline
Micro avg & 1.00 & 1.00 & 1.00 & 25 \\
\hline
Macro avg & 1.00 & 1.00 & 1.00 & 25 \\
\hline
Weighted avg & 1.00 & 1.00 & 1.00 & 25 \\
\hline
\end{tabular}
\caption{\label{tab:example4}Perfect identification of Acute Myeloid Leukemia (LAML) by GNB, SVM, DCT, RFC, LGR, KNN(K=3) all with 100\% sensitivity, specificity and precision.}
\end{table}

\begin{table}[ht]
\centering
\begin{tabular}{|l|l|l|l|l|}
\hline
ML / PAAD & Precision & Recall & F1-score & Support \\
\hline
Class 0 & 1.00 & 1.00 & 1.00 & 19 \\
\hline
Class 1 & 1.00 & 1.00 & 1.00 & 16 \\
\hline
Micro avg & 1.00 & 1.00 & 1.00 & 35 \\
\hline
Macro avg & 1.00 & 1.00 & 1.00 & 35 \\
\hline
Weighted avg & 1.00 & 1.00 & 1.00 & 35 \\
\hline
\end{tabular}
\caption{\label{tab:example5}Perfect identification of Pancreatic Adenocarcinoma by GNB, SVM, DCT, RFC, LGR, KNN(K=3) all with 100\% sensitivity, specificity and precision.}
\end{table}

\begin{table}[ht]
\centering
\begin{tabular}{|l|l|l|l|l|}
\hline
ML / TGCT & Precision & Recall & F1-score & Support \\
\hline
Class 0 & 1.00 & 1.00 & 1.00 & 22 \\
\hline
Class 1 & 1.00 & 1.00 & 1.00 & 8 \\
\hline
Micro avg & 1.00 & 1.00 & 1.00 & 30 \\
\hline
Macro avg & 1.00 & 1.00 & 1.00 & 30 \\
\hline
Weighted avg & 1.00 & 1.00 & 1.00 & 30 \\
\hline
\end{tabular}
\caption{\label{tab:example6}Perfect identification of Testicular Germ Cell Tumors by GNB, SVM, DCT, RFC, LGR, KNN(K=3) all with 100\% sensitivity, specificity and precision.}
\end{table}

\begin{table}[ht]
\centering
\begin{tabular}{|l|l|l|l|l|}
\hline
ML / ACC & Precision & Recall & F1-score & Support \\
\hline
Class 0 & 1.00 & 1.00 & 1.00 & 17 \\
\hline
Class 1 & 1.00 & 1.00 & 1.00 & 4 \\
\hline
Micro avg & 1.00 & 1.00 & 1.00 & 21 \\
\hline
Macro avg & 1.00 & 1.00 & 1.00 & 21 \\
\hline
Weighted avg & 1.00 & 1.00 & 1.00 & 21 \\
\hline
\end{tabular}
\caption{\label{tab:example7}Perfect identification of Adrenal Gland / Adrenocortical Carcinoma by GNB, SVM, RFC, LGR, KNN(K=3) all with 100\% sensitivity, specificity and precision.}
\end{table}

\begin{table}[ht]
\centering
\begin{tabular}{|l|l|l|l|l|}
\hline
DCT /ACC & Precision & Recall & F1-score & Support \\
\hline
Class 0 & 1.00 & 0.94 & 0.97 & 17 \\
\hline
Class 1 & 0.80 & 1.00 & 0.89 & 4 \\
\hline
Micro avg & 0.95 & 0.95 & 0.95 & 21 \\
\hline
Macro avg & 0.90 & 0.97 & 0.93 & 21 \\
\hline
Weighted avg & 0.95 & 0.95 & 0.95 & 21 \\
\hline
\end{tabular}
\caption{\label{tab:example8}Imperfect identification of Adrenal Gland / Adrenocortical Carcinoma by Decision Tree Classifier with 95\% accuracy.}
\end{table}

\begin{table}[ht]
\centering
\begin{tabular}{|l|l|l|l|l|}
\hline
GNB / Breast & Precision & Recall & F1-score & Support \\
\hline
Class 0 & 1.00 & 1.00 & 1.00 & 22 \\
\hline
Class 1 & 1.00 & 1.00 & 1.00 & 105 \\
\hline
Micro avg & 1.00 & 1.00 & 1.00 & 127 \\
\hline
Macro avg & 1.00 & 1.00 & 1.00 & 127 \\
\hline
Weighted avg & 1.00 & 1.00 & 1.00 & 127 \\
\hline
\end{tabular}
\caption{\label{tab:example9}Perfect identification of Breast invasive Carcinoma by GNB, SVM, RFC, LGR, KNN(K=3) all with 100\% sensitivity, specificity and precision.}
\end{table}

\begin{table}[ht]
\centering
\begin{tabular}{|l|l|l|l|l|}
\hline
ML / PRAD & Precision & Recall & F1-score & Support \\
\hline
Class 0 & 1.00 & 1.00 & 1.00 & 9 \\
\hline
Class 1 & 1.00 & 1.00 & 1.00 & 51 \\
\hline
Micro avg & 1.00 & 1.00 & 1.00 & 60 \\
\hline
Macro avg & 1.00 & 1.00 & 1.00 & 60 \\
\hline
Weighted avg & 1.00 & 1.00 & 1.00 & 60 \\
\hline
\end{tabular}
\caption{\label{tab:example10}Perfect identification of Prostate Adenocarcinoma by SVM, DCT, RFC and LGR all with 100\% sensitivity, specificity and precision.}
\end{table}

\begin{table}[ht]
\centering
\begin{tabular}{|l|l|l|l|l|}
\hline
ML / SKCM & Precision & Recall & F1-score & Support \\
\hline
Class 0 & 1.00 & 1.00 & 1.00 & 85 \\
\hline
Class 1 & 1.00 & 1.00 & 1.00 & 7 \\
\hline
Micro avg & 1.00 & 1.00 & 1.00 & 92 \\
\hline
Macro avg & 1.00 & 1.00 & 1.00 & 92 \\
\hline
Weighted avg & 1.00 & 1.00 & 1.00 & 92 \\
\hline
\end{tabular}
\caption{\label{tab:example11}Perfect identification of Skin Cutaneous Melanoma by GNB, SVM, DCT, RFC and LGR all with 100\% sensitivity, specificity and precision.}
\end{table}

\begin{table}[ht]
\centering
\begin{tabular}{|l|l|l|l|l|}
\hline
ML / STAD & Precision & Recall & F1-score & Support \\
\hline
Class 0 & 1.00 & 1.00 & 1.00 & 24 \\
\hline
Class 1 & 1.00 & 1.00 & 1.00 & 35 \\
\hline
Micro avg & 1.00 & 1.00 & 1.00 & 59 \\
\hline
Macro avg & 1.00 & 1.00 & 1.00 & 59 \\
\hline
Weighted avg & 1.00 & 1.00 & 1.00 & 59 \\
\hline
\end{tabular}
\caption{\label{tab:example12}Perfect identification of Stomach Adenocarcinoma by SVM, RFC and LGR all with 100\% sensitivity, specificity and precision.}
\end{table}

\begin{table}[ht]
\centering
\begin{tabular}{|l|l|l|l|l|}
\hline
ML / THCA & Precision & Recall & F1-score & Support \\
\hline
Class 0 & 1.00 & 1.00 & 1.00 & 29 \\
\hline
Class 1 & 1.00 & 1.00 & 1.00 & 50 \\
\hline
Micro avg & 1.00 & 1.00 & 1.00 & 79 \\
\hline
Macro avg & 1.00 & 1.00 & 1.00 & 79 \\
\hline
Weighted avg & 1.00 & 1.00 & 1.00 & 79  \\
\hline
\end{tabular}
\caption{\label{tab:example13}Perfect identification of Skin Cutaneous Melanoma by GNB, SVM, DCT, RFC and LGR all with 100\% sensitivity, specificity and precision.}
\end{table}

\begin{table}[ht]
\centering
\begin{tabular}{|l|l|l|l|l|}
\hline
ML / UCEC & Precision & Recall & F1-score & Support \\
\hline
Class 0 & 1.00 & 1.00 & 1.00 & 7 \\
\hline
Class 1 & 1.00 & 1.00 & 1.00 & 18 \\
\hline
Micro avg & 1.00 & 1.00 & 1.00 & 25 \\
\hline
Macro avg & 1.00 & 1.00 & 1.00 & 25 \\
\hline
Weighted avg & 1.00 & 1.00 & 1.00 & 25 \\
\hline
\end{tabular}
\caption{\label{tab:example14}Perfect identification of Uterine Corpus Endometrial Carcinoma (UCEC) by GNB, SVM, DCT, RFC, LGR, KNN(K=3) all with 100\% sensitivity, specificity and precision.}
\end{table}

\begin{table}[ht]
\centering
\begin{tabular}{|l|l|l|l|l|}
\hline
ML / OV & Precision & Recall & F1-score & Support \\
\hline
Class 0 & 1.00 & 1.00 & 1.00 & 3 \\
\hline
Class 1 & 1.00 & 1.00 & 1.00 & 48 \\
\hline
Micro avg & 1.00 & 1.00 & 1.00 & 51 \\
\hline
Macro avg & 1.00 & 1.00 & 1.00 & 51 \\
\hline
Weighted avg & 1.00 & 1.00 & 1.00 & 51 \\
\hline
\end{tabular}
\caption{\label{tab:example15}Perfect identification of Ovarian serous cystadenocarcinoma (OV) by GNB, SVM, DCT, RFC, LGR, KNN(K=3) all with 100\% sensitivity, specificity and precision.}
\end{table}

\begin{table}[ht]
\centering
\begin{tabular}{|l|l|l|l|l|}
\hline
ML / READ & Precision & Recall & F1-score & Support \\
\hline
Class 0 & 1.00 & 1.00 & 1.00 & 28 \\
\hline
Class 1 & 1.00 & 1.00 & 1.00 & 12 \\
\hline
Micro avg & 1.00 & 1.00 & 1.00 & 40 \\
\hline
Macro avg & 1.00 & 1.00 & 1.00 & 40 \\
\hline
Weighted avg & 1.00 & 1.00 & 1.00 & 40 \\
\hline
\end{tabular}
\caption{\label{tab:example16}Perfect identification of Rectum adenocarcinoma (READ) by GNB, SVM, DCT, RFC, LGR all with 100\% sensitivity, specificity and precision.}
\end{table}

\begin{table}[ht]
\centering
\begin{tabular}{|l|l|l|l|l|}
\hline
ML / READ & Precision & Recall & F1-score & Support \\
\hline
Class 0 & 1.00 & 1.00 & 1.00 & 1 \\
\hline
Class 1 & 1.00 & 1.00 & 1.00 & 55 \\
\hline
Micro avg & 1.00 & 1.00 & 1.00 & 56 \\
\hline
Macro avg & 1.00 & 1.00 & 1.00 & 56 \\
\hline
Weighted avg & 1.00 & 1.00 & 1.00 & 56 \\
\hline
\end{tabular}
\caption{\label{tab:example17}Perfect identification of Kidney cancer (KIRC) by SVM and LGR all with 100\% sensitivity, specificity and precision.}
\end{table}

\begin{table}[ht]
\centering
\begin{tabular}{|l|l|l|l|l|}
\hline
ML / READ & Precision & Recall & F1-score & Support \\
\hline
Class 0 & 1.00 & 1.00 & 1.00 & 8 \\
\hline
Class 1 & 1.00 & 1.00 & 1.00 & 160 \\
\hline
Micro avg & 1.00 & 1.00 & 1.00 & 168 \\
\hline
Macro avg & 1.00 & 1.00 & 1.00 & 168 \\
\hline
Weighted avg & 1.00 & 1.00 & 1.00 & 168 \\
\hline
\end{tabular}
\caption{\label{tab:example18}Perfect identification of Kidney cancer (KIRC) by RFC with 100\% sensitivity, specificity and precision when training with 70\% of total data. In this case, GNB, SVM, DCT and LGR returned 99\% accuracy.}
\end{table}

\begin{table}[ht]
\centering
\begin{tabular}{|l|l|l|l|l|}
\hline
ML / GBM & Precision & Recall & F1-score & Support \\
\hline
Class 0 & 1.00 & 1.00 & 1.00 & 118 \\
\hline
Class 1 & 1.00 & 1.00 & 1.00 & 13 \\
\hline
Micro avg & 1.00 & 1.00 & 1.00 & 131 \\
\hline
Macro avg & 1.00 & 1.00 & 1.00 & 131 \\
\hline
Weighted avg & 1.00 & 1.00 & 1.00 & 131 \\
\hline
\end{tabular}
\caption{\label{tab:example19}Perfect identification of Brain GBM Glioblastoma multiforme (GBM) by RFC, LGR, KNN(K=3) all with 100\% sensitivity, specificity and precision. GNB, SVM, DCT resulted in 99\% accuracy and KMeans 71\%; therefore were not shown in this table.}
\end{table}

\begin{table}[ht]
\centering
\begin{tabular}{|l|l|l|l|l|}
\hline
ML / LUAD & Precision & Recall & F1-score & Support \\
\hline
Class 0 & 1.00 & 1.00 & 1.00 & 20 \\
\hline
Class 1 & 1.00 & 1.00 & 1.00 & 61 \\
\hline
Micro avg & 1.00 & 1.00 & 1.00 & 81 \\
\hline
Macro avg & 1.00 & 1.00 & 1.00 & 81 \\
\hline
Weighted avg & 1.00 & 1.00 & 1.00 & 81 \\
\hline
\end{tabular}
\caption{\label{tab:example20}Perfect identification of Lung Adenocarcinoma (LUAD) by GNB, SVM, RFC, LGR all with 100\% sensitivity, specificity and precision.}
\end{table}

\begin{table}[ht]
\centering
\begin{tabular}{|l|l|l|l|l|}
\hline
ML / LUSC & Precision & Recall & F1-score & Support \\
\hline
Class 0 & 1.00 & 1.00 & 1.00 & 32 \\
\hline
Class 1 & 1.00 & 1.00 & 1.00 & 47 \\
\hline
Micro avg & 1.00 & 1.00 & 1.00 & 79 \\
\hline
Macro avg & 1.00 & 1.00 & 1.00 & 79 \\
\hline
Weighted avg & 1.00 & 1.00 & 1.00 & 79 \\
\hline
\end{tabular}
\caption{\label{tab:example21}Perfect identification of Lung squamous cell carcinoma (LUSC) by GNB, SVM, RFC, LGR all with 100\% sensitivity, specificity and precision.}
\end{table}

\begin{table}[ht]
\centering
\begin{tabular}{|l|l|l|l|l|}
\hline
DCT / Lung and (LUAD and LUSC) & Precision & Recall & F1-score & Support \\
\hline
Class 0 & 1.00 & 1.00 & 1.00 & 24 \\
\hline
Class 1 & 1.00 & 1.00 & 1.00 & 106 \\
\hline
Micro avg & 1.00 & 1.00 & 1.00 & 130 \\
\hline
Macro avg & 1.00 & 1.00 & 1.00 & 130 \\
\hline
Weighted avg & 1.00 & 1.00 & 1.00 & 130 \\
\hline
\end{tabular}
\caption{\label{tab:example22}Perfect identification of Lung cancers (both LUAD and LUSC together) by DCT with 100\% sensitivity, specificity and precision.}
\end{table}

\begin{table}[ht]
\centering
\begin{tabular}{|l|l|l|l|l|}
\hline
ML / Lung and (LUAD and LUSC) & Precision & Recall & F1-score & Support \\
\hline
Class 0 & 1.00 & 0.92 & 0.96 & 24 \\
\hline
Class 1 & 0.98 & 1.00 & 0.99 & 106 \\
\hline
Micro avg & 0.98 & 0.98 & 0.98 & 130 \\
\hline
Macro avg & 0.99 & 0.96 & 0.97 & 130 \\
\hline
Weighted avg & 0.98 & 0.98 & 0.98 & 130 \\
\hline
\end{tabular}
\caption{\label{tab:example23}Imperfect results of identification of Lung cancers (both LUAD and LUSC together) and Lung healthy tissue by SVM, RFC, LGR all with 98\% accuracy (F-score).}
\end{table}

% \caption{\label{tab:example}Perfect identification of Kidney renal clear cell carcinoma (KIRC) by GNB, SVM, DCT, RFC, LGR and K=5NN.}
% \end{table}

% \begin{table}[ht]
% \centering
% \begin{tabular}{|l|l|l|l|l|}
% \hline
% KNN / CESC & Precision & Recall & F1-score & Support \\
% \hline
% Class 0 & 0.75 & 1.00 & 0.86 & 17 \\
% \hline
% Class 1 & 1.00 & 0.99 & 0.99 & 4 \\
% \hline
% Micro avg & 0.99 & 0.99 & 0.99 & 21 \\
% \hline
% Macro avg & 0.88 & 0.99 & 0.93 & 21 \\
% \hline
% Weighted avg & 0.99 & 0.99 & 0.99 & 21 \\
% \hline
% \end{tabular}
% \caption{\label{tab:example}Imperfect identification of Endocervical Adenocarcinoma by K=3Nearest Neighbors.}
% \end{table}

\begin{table}[ht]
\centering
\begin{tabular}{|l|l|l|l|l|}
\hline
SVM, DCT and LGR / ESCA & Precision & Recall & F1-score & Support \\
\hline
Class 0 & 1.00 & 1.00 & 1.00 & 69 \\
\hline
Class 1 & 1.00 & 1.00 & 1.00 & 15 \\
\hline
Micro avg & 1.00 & 1.00 & 1.00 & 84 \\
\hline
Macro avg & 1.00 & 1.00 & 1.00 & 84 \\
\hline
Weighted avg & 1.00 & 1.00 & 1.00 & 84 \\
\hline
\end{tabular}
\caption{\label{tab:example24}Perfect identification of Esophageal Carcinoma by SVM, DCT and LGR all with 100\% sensitivity, specificity and precision.}
\end{table}

\begin{table}[ht]
\centering
\begin{tabular}{|l|l|l|l|l|}
\hline
SVM, RFC and LGR / LIHC & Precision & Recall & F1-score & Support \\
\hline
Class 0 & 1.00 & 1.00 & 1.00 & 69 \\
\hline
Class 1 & 1.00 & 1.00 & 1.00 & 15 \\
\hline
Micro avg & 1.00 & 1.00 & 1.00 & 84 \\
\hline
Macro avg & 1.00 & 1.00 & 1.00 & 84 \\
\hline
Weighted avg & 1.00 & 1.00 & 1.00 & 84 \\
\hline
\end{tabular}
\caption{\label{tab:example25}Perfect identification of Liver and Hepatocellular Carcinoma by SVM, DCT and LGR all with 100\% sensitivity, specificity and precision.} 
\end{table}

\begin{table}[ht]
\centering
\begin{tabular}{|l|l|l|l|l|}
\hline
RFC / BLCA & Precision & Recall & F1-score & Support \\
\hline
Class 0 & 1.00 & 1.00 & 1.00 & 3 \\
\hline
Class 1 & 1.00 & 1.00 & 1.00 & 247 \\
\hline
Micro avg & 1.00 & 1.00 & 1.00 & 250 \\
\hline
Macro avg & 1.00 & 1.00 & 1.00 & 250 \\
\hline
Weighted avg & 1.00 & 1.00 & 1.00 & 250 \\
\hline
\end{tabular}
\caption{\label{tab:example26}Perfect identification of Bladder Urothelial Carcinoma by RFC. This particular result is achieved for train-test slpit on 40\% and 60\% due to low number of healthy bladder samples. All other 5 classifeirs also show excellent performance on separation between high-dimensional WES data of cancerous and noncancerous cells despite having little data to train.}
\end{table}

\begin{table}[ht]
\centering
\begin{tabular}{|l|l|l|l|l|}
\hline
GNB / LGG & Precision & Recall & F1-score & Support \\
\hline
Class 0 & 0.98 & 0.99 & 0.99 & 106 \\
\hline
Class 1 & 0.98 & 0.97 & 0.97 & 60 \\
\hline
Micro avg & 0.98 & 0.98 & 0.98 & 166 \\
\hline
Macro avg & 0.98 & 0.98 & 0.98 & 166 \\
\hline
Weighted avg & 0.98 & 0.98 & 0.98 & 166 \\
\hline
\end{tabular}
\caption{\label{tab:example27}Naive Bayes Classification Report on Brain Cancer.}
\end{table}

\begin{table}[ht]
\centering
\begin{tabular}{|l|l|l|l|l|}
\hline
SVM / LGG & Precision & Recall & F1-score & Support \\
\hline
Class 0 & 1.00 & 0.99 & 1.00 & 106 \\
\hline
Class 1 & 0.98 & 1.00 & 0.99 & 60 \\
\hline
Micro avg & 0.99 & 0.99 & 0.99 & 166 \\
\hline
Macro avg & 0.99 & 1.00 & 0.99 & 166 \\
\hline
Weighted avg & 0.99 & 0.99 & 0.99 & 166 \\
\hline
\end{tabular}
\caption{\label{tab:example28}SVM Classification Report on Brain Cancer.}
\end{table}

\begin{table}[ht]
\centering
\begin{tabular}{|l|l|l|l|l|}
\hline
DCT / LGG & Precision & Recall & F1-score & Support \\
\hline
Class 0 & 1.00 & 1.00 & 1.00 & 106 \\
\hline
Class 1 & 1.00 & 1.00 & 1.00 & 60 \\
\hline
Micro avg & 1.00 & 1.00 & 1.00 & 166 \\
\hline
Macro avg & 1.00 & 1.00 & 1.00 & 166 \\
\hline
Weighted avg & 1.00 & 1.00 & 1.00 & 166 \\
\hline
\end{tabular}
\caption{\label{tab:example29}Decision Tree Classification Report on Brain Cancer.}
\end{table}

\begin{table}[ht]
\centering
\begin{tabular}{|l|l|l|l|l|}
\hline
RFC / LGG & Precision & Recall & F1-score & Support \\
\hline
Class 0 & 0.98 & 1.00 & 0.99 & 106 \\
\hline
Class 1 & 1.00 & 0.97 & 0.98 & 60 \\
\hline
Micro avg & 0.99 & 0.99 & 0.99 & 166 \\
\hline
Macro avg & 0.99 & 0.98 & 0.99 & 166 \\
\hline
Weighted avg & 0.99 & 0.99 & 0.99 & 166 \\
\hline
\end{tabular}
\caption{\label{tab:example30}Random Forest Classification Report on Brain Cancer.}
\end{table}

\begin{table}[ht]
\centering
\begin{tabular}{|l|l|l|l|l|}
\hline
LGR / LGG & Precision & Recall & F1-score & Support \\
\hline
Class 0 & 1.00 & 0.99 & 1.00 & 106 \\
\hline
Class 1 & 0.98 & 1.00 & 0.99 & 60 \\
\hline
Micro avg & 0.99 & 0.99 & 0.99 & 166 \\
\hline
Macro avg & 0.99 & 1.00 & 0.99 & 166 \\
\hline
Weighted avg & 0.99 & 0.99 & 0.99 & 166 \\
\hline
\end{tabular}
\caption{\label{tab:example31}Logistic Regression Classification Report on Brain Cancer.}
\end{table}

\begin{table}[ht]
\centering
\begin{tabular}{|l|l|l|l|l|}
\hline
KNN / LGG & Precision & Recall & F1-score & Support \\
\hline
Class 0 & 1.00 & 1.00 & 1.00 & 106 \\
\hline
Class 1 & 1.00 & 1.00 & 1.00 & 60 \\
\hline
Micro avg & 1.00 & 1.00 & 1.00 & 166 \\
\hline
Macro avg & 1.00 & 1.00 & 1.00 & 166 \\
\hline
Weighted avg & 1.00 & 1.00 & 1.00 & 166 \\
\hline
\end{tabular}
\caption{\label{tab:example32} K=3 Nearest Neighbors and K-Means Classification Report on Brain Cancer.}
\end{table}

% \begin{table}[ht]
% \centering
% \begin{tabular}{|l|l|l|l|l|}
% \hline
% KNN / Breast & Precision & Recall & F1-score & Support \\
% \hline
% Class 0 & 0.91 & 0.95 & 0.93 & 22 \\
% \hline
% Class 1 & 0.99 & 0.98 & 0.99 & 105 \\
% \hline
% Micro avg & 0.98 & 0.98 & 0.98 & 127 \\
% \hline
% Macro avg & 0.95 & 0.97 & 0.96 & 127 \\
% \hline
% Weighted avg & 0.98 & 0.98 & 0.98 & 127 \\
% \hline
% \end{tabular}
% \caption{\label{tab:example}Legend (350 words max). Example legend text.}
% \end{table}

% \begin{table}[ht]
% \centering
% \begin{tabular}{|l|l|l|l|l|}
% \hline
% DCT / Breast & Precision & Recall & F1-score & Support \\
% \hline
% Class 0 & 1.00 & 0.95 & 0.98 & 22 \\
% \hline
% Class 1 & 0.99 & 1.00 & 1.00 & 105 \\
% \hline
% Micro avg & 0.99 & 0.99 & 0.99 & 127 \\
% \hline
% Macro avg & 1.00 & 0.98 & 0.99 & 127 \\
% \hline
% Weighted avg & 0.99 & 0.99 & 0.99 & 127 \\
% \hline
% \end{tabular}
% \caption{\label{tab:example}Legend (350 words max). Example legend text.}
% \end{table}

\begin{table}[ht]
\centering
\begin{tabular}{|l|l|l|l|l|}
\hline
GNB and DCT / Colon & Precision & Recall & F1-score & Support \\
\hline
Class 0 & 1.00 & 1.00 & 1.00 & 30 \\
\hline
Class 1 & 1.00 & 1.00 & 1.00 & 30 \\
\hline
Micro avg & 1.00 & 1.00 & 1.00 & 60 \\
\hline
Macro avg & 1.00 & 1.00 & 1.00 & 60 \\
\hline
Weighted avg & 1.00 & 1.00 & 1.00 & 60 \\
\hline
\end{tabular}
\caption{\label{tab:example33}Legend (350 words max). Example legend text.}
\end{table}

\begin{table}[ht]
\centering
\begin{tabular}{|l|l|l|l|l|}
\hline
SVM / Colon & Precision & Recall & F1-score & Support \\
\hline
Class 0 & 0.91 & 0.95 & 0.93 & 30 \\
\hline
Class 1 & 0.99 & 0.98 & 0.99 & 30 \\
\hline
Micro avg & 0.98 & 0.98 & 0.98 & 60 \\
\hline
Macro avg & 0.95 & 0.97 & 0.96 & 60 \\
\hline
Weighted avg & 0.98 & 0.98 & 0.98 & 60 \\
\hline
\end{tabular}
\caption{\label{tab:example34}Legend (350 words max). Example legend text.}
\end{table}

\begin{table}[ht]
\centering
\begin{tabular}{|l|l|l|l|l|}
\hline
RFC / Colon & Precision & Recall & F1-score & Support \\
\hline
Class 0 & 0.97 & 1.00 & 0.98 & 30 \\
\hline
Class 1 & 1.00 & 0.97 & 0.98 & 30 \\
\hline
Micro avg & 0.98 & 0.98 & 0.98 & 60 \\
\hline
Macro avg & 0.98 & 0.98 & 0.98 & 60 \\
\hline
Weighted avg & 0.98 & 0.98 & 0.98 & 60 \\
\hline
\end{tabular}
\caption{\label{tab:example35}Legend (350 words max). Example legend text.}
\end{table}

\begin{table}[ht]
\centering
\begin{tabular}{|l|l|l|l|l|}
\hline
LGR / Colon & Precision & Recall & F1-score & Support \\
\hline
Class 0 & 1.00 & 0.97 & 0.98 & 30 \\
\hline
Class 1 & 0.97 & 1.00 & 0.98 & 30 \\
\hline
Micro avg & 0.98 & 0.98 & 0.98 & 60 \\
\hline
Macro avg & 0.98 & 0.98 & 0.98 & 60 \\
\hline
Weighted avg & 0.98 & 0.98 & 0.98 & 60 \\
\hline
\end{tabular}
\caption{\label{tab:example36}Legend (350 words max). Example legend text.}
\end{table}

\begin{table}[ht]
\centering
\begin{tabular}{|l|l|l|l|l|}
\hline
KNN / Colon & Precision & Recall & F1-score & Support \\
\hline
Class 0 & 1.00 & 0.93 & 0.97 & 30 \\
\hline
Class 1 & 0.94 & 1.00 & 0.97 & 30 \\
\hline
Micro avg &  0.97 & 0.97 & 0.97 & 60 \\
\hline
Macro avg &  0.97 & 0.97 & 0.97 & 60 \\
\hline
Weighted avg & 0.97 & 0.97 & 0.97 & 60 \\
\hline
\end{tabular}
\caption{\label{tab:example37}Legend (350 words max). Example legend text.}
\end{table}

\begin{table}[ht]
\centering
\begin{tabular}{|l|l|l|l|l|}
\hline
GNB and RFC / ESCA & Precision & Recall & F1-score & Support \\
\hline
Class 0 & 0.99 & 1.00 & 0.99 & 69 \\
\hline
Class 1 & 1.00 & 0.93 & 0.97 & 15 \\
\hline
Micro avg & 0.99 & 0.99 & 0.99 & 84 \\
\hline
Macro avg & 0.99 & 0.97 & 0.98 & 84 \\
\hline
Weighted avg & 0.99 & 0.99 & 0.99 & 84 \\
\hline
\end{tabular}
\caption{\label{tab:example38} Naive Bayes and Random Forest Classification Report on Esophagus and Esophageal Carcinoma}
\end{table}

\begin{table}[ht]
\centering
\begin{tabular}{|l|l|l|l|l|}
\hline
GNB / LIHC & Precision & Recall & F1-score & Support \\
\hline
Class 0 & 0.94 & 1.00 & 0.97 & 15 \\
\hline
Class 1 & 1.00 & 0.97 & 0.98 & 33 \\
\hline
Micro avg & 0.98 & 0.98 & 0.98 & 48 \\
\hline
Macro avg & 0.97 & 0.98 & 0.98 & 48 \\
\hline
Weighted avg & 0.98 & 0.98 & 0.98 & 48 \\
\hline
\end{tabular}
\caption{\label{tab:example39}Performance of GNB on Liver LIHC classification with 0.1 data as testing data and random seed equal to 42 instead of 0.}
\end{table}

\begin{table}[ht]
\centering
\begin{tabular}{|l|l|l|l|l|}
\hline
ML / LIHC & Precision & Recall & F1-score & Support \\
\hline
Class 0 & 1.00 & 1.00 & 1.00 & 15 \\
\hline
Class 1 & 1.00 & 1.00 & 1.00 & 33 \\
\hline
Micro avg & 1.00 & 1.00 & 1.00 & 48 \\
\hline
Macro avg & 1.00 & 1.00 & 1.00 & 48 \\
\hline
Weighted avg & 1.00 & 1.00 & 1.00 & 48 \\
\hline
\end{tabular}
\caption{\label{tab:example40}Perfect identification of Liver LIHC cancer by SVM, DCT, RFC, LGR and K=3NN with 0.1 data as testing data and random seed equal to 42 instead of 0. KMeans clustering was done with accuracy of 94\%}
\end{table}

\begin{table}[ht]
\centering
\begin{tabular}{|l|l|l|l|l|}
\hline
SVM with rbf kernel / LUSC & Precision & Recall & F1-score & Support \\
\hline
Class 0 & 0.00 & 0.00 & 0.00 & 32 \\
\hline
Class 1 & 0.59 & 1.00 & 0.75 & 47 \\
\hline
Micro avg & 0.59 & 0.59 & 0.59 & 79 \\
\hline
Macro avg & 0.30 & 0.50 & 0.37 & 79 \\
\hline
Weighted avg & 0.35 & 0.59 & 0.44 & 79 \\
\hline
\end{tabular}
\caption{\label{tab:example41} The performance of SVM classifier on Lung and LUSC (Lung Squamous Cell Carcinoma), when SVM using rbf as kernel with no correct detection of healthy cells, instead of linear kernel that resulted in 100\% accurate classification of both classes, shows that the nature of separation between healthy and cancerous cells is linear.}
\end{table}

\begin{table}[ht]
\centering
\begin{tabular}{|l|l|l|l|l|}
\hline
GNB, SVM, RFC, LGR / LUSC & Precision & Recall & F1-score & Support \\
\hline
Class 0 & 1.00 & 1.00 & 1.00 & 32 \\
\hline
Class 1 & 1.00 & 1.00 & 1.00 & 47 \\
\hline
Micro avg & 1.00 & 1.00 & 1.00 & 79 \\
\hline
Macro avg & 1.00 & 1.00 & 1.00 & 79 \\
\hline
Weighted avg & 1.00 & 1.00 & 1.00 & 79 \\
\hline
\end{tabular}
\caption{\label{tab:example42}Perfect identification of Lung squamous cell carcinoma by GNB, SVM, RFC, LGR all with 100\% sensitivity, specificity and precision. Even KNN(K=3) returned the same perfect result when using random seed 42 instead of 0 that returned 0.98 that is also for DCT.}
\end{table}

\begin{table}[ht]
\centering
\begin{tabular}{|l|l|l|l|l|l|l|l|}
\hline
Accuracy based on training data &  GNB & SVM & DCT & RFC & LGR & KNN & KMeans unsupervised clustering \\
\hline
0.1 & 0.96 & 0.94 & 0.96 & 0.98 & 0.96 & 0.94 & 0.92 \\
\hline
0.3 & 0.98 & 0.99 & 0.97 & 0.99 & 0.99 & 0.95 & 0.93 \\
\hline
0.5 & 0.99 & 0.98 & 0.96 & 0.99 & 0.99 & 0.95 & 0.94 \\
\hline
0.7 & 0.99 & 0.99 & 0.99 & 1.0 & 1.0 & 0.94 & 0.93 \\
\hline
0.9 & 0.92 & 1.0 & 0.98 & 1.0 & 1.0 & 0.96 & 0.92 \\
\hline
\end{tabular}
\caption{\label{tab:example43}The effect of different portion of LIHC data as training data volume on F-score/accuracy of different classifiers.}
\end{table}

\begin{table}[ht]
\centering
\begin{tabular}{|l|l|l|l|}
\hline
KNN F-score on & Liver and LIHC & Lung and LUSC & BLCA and LUSC cancers together \\
\hline
K = 1 & 0.94 & 1.0 & 0.85 \\
\hline
K = 3 & 0.96 & 0.98 & 0.89 \\
\hline
K = 5 & 0.96 & 0.99 & 0.91 \\
\hline
K = 7 & 0.96 &0.99 & 0.91 \\
\hline
K = 13 & 0.96 & 0.99 & 0.85 \\
\hline
K = 39 & 0.96 & 0.96 & 0.84 \\
\hline
\end{tabular}
\caption{\label{tab:example44}Performance analysis by F-score accuracy of KNN on Liver LIHC classification with 0.1 data as testing data (10 Class 0 and 38 Class 1 samples) and random seed equal to 0 but with different values for K.}
\end{table}

\begin{table}[ht]
\centering
\begin{tabular}{|l|l|l|l|l|l|l|l|}
\hline
ML / 2 Cancers &  GNB & SVM & DCT & RFC & LGR & KNN & KMeans unsupervised clustering \\
\hline
Lung LUAD and LUSC & 0.94 & 0.90 & 0.94 & 0.95 & 0.91 & 0.88 & 0.38=0.62 \\
\hline
Brain GBM and LGG & 0.94 & 0.98 & 0.98 & 1.0 & 0.94 & 0.92 & 0.75 \\
\hline
Colon COAD and READ & 0.61 & 0.68 & 0.74 & 0.76 & 0.68 & 0.74 & 0.71 \\
\hline
Kidney KIRC and KIRP & 0.91 & 0.91 & 0.89 & 0.93 & 0.91 & 0.88 & 0.57 \\
\hline
Lung LUSC and Bladder BLCA & 0.97 & 0.98 & 0.98 & 0.99 & 1.0 & 0.91 & 0.68 \\
\hline
\end{tabular}
\caption{\label{tab:example45}The F-score/accuracy performance of different ML classifiers on 2 types of cancerous samples.}
\end{table}

%Figures and tables can be referenced in LaTeX using the ref command, e.g. Figure \ref{fig:stream} and Table \ref{tab:example}.


\begin{thebibliography}{}
%\noindent 
\bibitem{b1} Siegel, Rebecca L., Kimberly D. Miller, and Ahmedin Jemal, Cancer Statistics, 2020, CA: A Cancer Journal for Clinicians 70, no. 1: 7-30,  (2020).
\bibitem{b2} Momenimovahed, Zohre, and Hamid Salehiniya, Epidemiological characteristics of and risk factors for breast cancer in the world, Breast Cancer: Targets and Therapy 11: 151, (2019).
\bibitem{b3} Tsuji, S., and H. Aburatani., Machine Learning Applications in Cancer Genome Medicine, Gan to kagaku ryoho. Cancer \& chemotherapy 46.3:: 423-426, (2019).
\bibitem{b4} Nik-Zainal Abidin, S., Memari, Y., \& Davies, H., Holistic cancer genome profiling for every patient, Swiss medical weekly, 150 w20158, 20158 https://doi.org/10.4414/smw.(2020).
\bibitem{b5} Asri, Hiba, et al., Using machine learning algorithms for breast cancer risk prediction and diagnosis, Procedia Computer Science 83: 1064-1069, (2016).
\bibitem{b6} Vamathevan, Jessica, et al. Applications of machine learning in drug discovery and development, Nature Reviews Drug Discovery, 18.6, 463-477, (2019). https://doi.org/10.1038/s41573-019-0024-5
\bibitem{b7} Streiner, David L., Clinical medicine and the legacy of the reverend Bayes, International journal of clinical practice 73.4:e13323, (2019).
\bibitem{b8} Myung, Jae., Tutorial on maximum likelihood estimation, journal of mathematical psychology, 47 (2003), 90100, (2002).
\bibitem{b9} Zhang, Harry., The optimality of naive Bayes, AA 1.2, 3 (2004).
\bibitem{b10} Pes, Barbara., Ensemble feature selection for high-dimensional data: a stability analysis across multiple domains, Neural Computing and Applications: 1-23, (2019).
\bibitem{b11} Sun, Yingshuai, et al., Identification of 12 cancer types through genome deep learning, Nature Scientific reports 9.1: 1-9 (2019).
\bibitem{b12} Abeel, Thomas, et al., Robust biomarker identification for cancer diagnosis with ensemble feature selection methods, Bioinformatics 26.3: 392-398, (2010).
\bibitem{b13} Sidney, S., Go, A. S., \& Rana, J. S., Transition From Heart Disease to Cancer as the Leading Cause of Death in the United States, Annals of internal medicine, 171(3): 225, (2019).
\bibitem{b14} Alabsi, A. M., Ali, R., Ali, A. M., Al-Dubai, S. A. R., Harun, H., Abu Kasim, N. H., \& Alsalahi, A., Apoptosis induction, cell cycle arrest and in vitro anticancer activity of gonothalamin in a cancer cell lines, Asian Pacific Journal of Cancer Prevention, 13(10), 5131-5136, (2012).

\bibitem{b15} Tomczak, K., Czerwinska, P., and Wiznerowicz, M., The Cancer Genome Atlas (TCGA): an immeasurable source of knowledge, Contemporary oncology, 19(1A), A68, (2015).
\bibitem{b16} Lonsdale, J., Thomas, J., Salvatore, M., Phillips, R., Lo, E., Shad, S., ... and Foster, B., The genotype-tissue expression (GTEx) project, Nature genetics, 45(6), 580-585, (2013).
\bibitem{b17} Cortes, C., and Vapnik, V., Support-vector networks, Machine learning, 20(3), 273-297, (1995).
\bibitem{b18} Boser, B. E., Guyon, I. M., and Vapnik, V. N., A training algorithm for optimal margin classifiers, In Proceedings of the fifth annual workshop on Computational learning theory (pp. 144-152), (1992).
\bibitem{b19} Ben-Hur, A., Horn, D., Siegelmann, H. T., and Vapnik, V., Support vector clustering, Journal of machine learning research, 2(Dec), 125-137, (2002).
\bibitem{b20} Noble, W. S., What is a support vector machine?, Nature biotechnology, 24(12), 1565-1567, (2006).
\bibitem{b21} Breiman, L., Friedman, J. H., Olshen, R. A., and Stone, C. J., Classification and regression trees, Statistics/probability series, (1984).
\bibitem{b22} Ho, T. K., Random decision forests, In Proceedings of 3rd international conference on document analysis and recognition (Vol. 1, pp. 278-282). IEEE, (1995).
\bibitem{b23} Ho, T. K., The random subspace method for constructing decision forests, IEEE transactions on pattern analysis and machine intelligence, 20(8), 832-844, (1998).
\bibitem{b24} Cox, D. R., The regression analysis of binary sequences. Journal of the Royal Statistical Society: Series B (Methodological), 20(2), 215-232, (1958).
\bibitem{b25} Fix, E., and Hodges, J. L., Discriminatory analysis, nonparametric discrimination, (1951).
\bibitem{b26} Peterson, L. E., K-nearest neighbor. Scholarpedia, 4(2), 1883, (2009).
\bibitem{b27} MacQueen, J., Some methods for classification and analysis of multivariate observations. In Proceedings of the fifth Berkeley symposium on mathematical statistics and probability, Vol. 1, No. 14, pp. 281-297, (1967).
\end{thebibliography}
\end{document}